\documentclass[12pt]{article}
\usepackage[dvips]{graphicx}

\def\lromn#1{\uppercase\expandafter{\romannumeral#1}}

\begin{document}
  
\begin{titlepage}

\begin{center}

\hfill KEK-TH-1171 \\
\hfill August 2007
%\hfill \today

\vspace{1cm}
{\large\bf Higgs Particle: The Origin of Mass}
\vspace{1.5cm}

{\bf Yasuhiro Okada}$^{(a,b)}$
\footnote{E-mail: yasuhiro.okada@kek.jp}
\vskip 0.2in
{\it
$^{(a)}${Theory Group, KEK, Oho 1-1 Tsukuba, 305-0801, Japan} \\
$^{(b)}${The Graduate University for Advanced Studies (Sokendai),
\\
Oho 1-1 Tsukuba, 305-0801, Japan}\\
}

\vskip 1in

\abstract{The Higgs particle is a new elementary particle predicted 
in the Standard Model of the elementary particle physics. 
It plays a special role in the theory of mass generation
of quarks, leptons, and gauge bosons. In this article, theoretical 
issues on the Higgs mechanism are first discussed, and then  
experimental prospects on the Higgs particle study
at the future collider experiments, LHC and ILC, are 
reviewed. The Higgs coupling determination is an essential step 
to establish the mass generation mechanism, which could 
lead to a deeper understanding of particle physics.
}

\end{center}
\end{titlepage}
\setcounter{footnote}{0}

\section{Introduction} 
%% No sections necessary for express letters, letters and short note
Current understanding of the elementary particle physics is based
on two important concepts, gauge invariance and spontaneous symmetry
breaking. Out of four fundamental interactions of Nature,
namely strong, weak and electromagnetic and gravity interactions,
three of them except for gravity are described on the same footing 
in terms of gauge theory. The gauge group corresponding to the strong 
interaction is $SU(3)$, and the weak and the electromagnetic interactions
arise from $SU(2)$ and $U(1)$ groups and are called the
electroweak interaction. Once quarks and leptons are assigned in
proper representations of the three gauge groups, all properties 
of the three fundamental interactions are determined from 
the requirement of gauge invariance.

For more than thirty years, high energy experiments have been
testing various aspects of gauge symmetry and have established 
gauge invariance as a fundamental principle of Nature. We have discovered 
gauge bosons mediating the three interactions, namely gluon for the strong interaction, W and Z bosons for the electroweak interaction. The couplings 
between quarks/leptons and gauge bosons have been precisely measured 
at the CERN LEP and SLAC SLC experiments, and we have confirmed 
the assignments of the gauge representations for quarks and leptons.

The gauge principle alone, however, cannot describe the known 
structure of the elementary particle physics. In the Standard Model of 
the elementary particle physics, all quarks, leptons and gauge bosons 
are first introduced as massless fields.
In order to generate masses for these particles, the 
$SU(2) \times U(1)$ symmetries have to be broken spontaneously.

Spontaneous symmetry breaking itself is not new for particle physics
\cite{Nambu:1961tp, Goldstone:1961eq}.
The theory of the strong interaction, QCD, possesses an approximate symmetry
among three light quarks called chiral symmetry. The vacuum of QCD
corresponds to a state where quark and anti-quark pair
is condensed, and the chiral symmetry is broken spontaneously.
As a consequence, pseudo scalar mesons such as pions and kaons
are light compared to the typical energy scale of the strong interaction
since they behave approximately as Nambu-Goldstone bosons, 
a characteristic signature of spontaneous symmetry breaking. 

In the case of the electroweak symmetry, it is 
shown theoretically that the Nambu-Goldstone bosons associated 
with spontaneous breakdown are absorbed by gauge bosons, providing 
the mass generation mechanism for gauge bosons (Higgs mechanism) 
\cite{Higgs:1964ia}.  
Although we are now quite sure that this is the mechanism for 
gauge boson mass generation, we know little about how the symmetry 
breaking occurs, or what is dynamics behind the Higgs 
mechanism. Clearly, we need a new interaction other than four 
known fundamental forces, but we do not know what it is.  
The goal of the Higgs physics is to answer this question.

In this article, I would like to explain what are theoretical issues 
of the Higgs sector, what is expected at the future collider
experiments, LHC and ILC, and what would be impacts of the Higgs physics
on a deeper understanding of the particle physics.

\section{Higgs boson in the Standard Model}
In the Standard Model, a single Higgs doublet field is included for
the symmetry breaking of the $SU(2)\times U(1)$ gauge groups. This
was introduced in S. Weinberg's 1967 paper 
"A Model of Leptons" \cite{Weinberg:1967tq}, and 
is the simplest possibility for generating the gauge boson masses.

The Higgs potential is given by
\begin{equation}
V(\Phi)=-\mu^2|\Phi|^2+\lambda |\Phi|^4,
\label{eq1}
\end{equation}
where the two component complex field is defined as
\begin{equation}
\Phi(x)=\left(
\begin{array}{c}
\phi(x)^+\\
\phi(x)^0
\end{array}
\right).
\end{equation}
In order for the stability of the vacuum the parameter $\lambda$ must 
be positive. The coefficient of the quadratic term, on the other hand, can be
either sign. In fact, if the sign is negative, namely $\mu^2>0$, the origin of
the potential is unstable, and the vacuum state corresponds to a non-zero
value of the $\Phi$ field. The states satisfying  
$|\phi^+|^2+|\phi^0|^2=\frac{\mu^2}{2\lambda}\equiv\frac{v^2}{2}$ are
degenerate minimum of the potential.  We can choose the vacuum expectation value 
in the $\phi^0$ direction, $<\phi^0>=\frac{v}{\sqrt{2}}$, and then there are
three massless modes corresponding to the flat directions of the potential
(Nambu-Goldstone modes). When the symmetry is a gauge symmetry,
these massless particles disappear from the physical spectrum, and become
longitudinal components of massive gauge bosons. This is seen most clearly
if we take the "Unitary gauge" where the Nambu-Goldstone modes are removed
by an appropriate gauge transformation. The kinetic term of the scalar field 
is defined as
$|D_{\mu}\Phi|^2=|(\partial_{\mu}+g\frac{\tau^a}{2}W_{\mu}^a+\frac{g'}{2}B_{\mu})\Phi|^2$, and the gauge boson mass terms are obtained by substituting
the vacuum expectation value into $\Phi(x)$.   

Mass terms of quarks and leptons are also generated through interactions 
with the Higgs field. This follows from the chiral structure of quarks and
leptons. Since the discovery of the parity violation in the weak 
interaction \cite{Lee:1956qn}, chiral projected fermions (Weyl fermions) 
instead of Dirac fermions have been considered as building 
blocks of a particle physics model. 
In particular, only left-handed quarks and leptons are assigned 
as $SU(2)$ doublets because the weak interaction has a  
V(vector)-A(axial vector) current structure.  Right-handed 
counter parts are singlet under the $SU(2)$ gauge group. This unbalance 
in the $SU(2)$ quantum number assignment forbids us to write direct mass 
terms for quarks and leptons: the only possible way to generate mass 
terms is to introduce Yukawa couplings with help of the $\Phi$ field 
such as $y_d\Phi^{\dagger}\bar{d}_R q_L$ where $q_L=(u_L,d_L)^T$. 
After replacing $\Phi(x)$ by its vacuum  expectation value, this term 
generates a mass of $y_d v/\sqrt{2}$ for down-type quarks. Similar mechanism
works for up-type quarks and charged leptons.

There is one important prediction of this model. Since we introduce
a two-component complex field and three real degrees of freedom are 
absorbed by gauge bosons, one scalar particle appears in the physical
spectrum, which is called the Higgs particle ($\equiv$ Higgs boson). 
In the Unitary gauge, interactions related to the Higgs boson
can be obtained by replacing $v$ with $v+H(x)$ in
the Lagrangian where $H(x)$ represents the Higgs boson. 
The mass of the Higgs boson is given by 
$m_h=\sqrt{2\lambda}v$, which means that the Higgs boson becomes heavier
if the Higgs self-coupling gets larger. In fact, this is a general 
property of the particle mass generation mechanism due to the  Higgs field: 
A stronger interaction leads to a heaver particle. The mass formula 
for the W, Z bosons, quarks, leptons and the Higgs boson at the lowest 
order approximation with respect to coupling perturbation (i.e. tree-revel)
are summarized in table \ref{t1}.  
\begin{table}[htbp]
 \caption{Mass formula for elementary particles. $g$, $g'$, $y_f$,
 and $\lambda$ are the $SU(2)$ and the $U(1)$ gauge coupling constants, the
 Yukawa coupling constant for a fermion $f$, and the Higgs 
 self-coupling constant.  }
 \label{t1}
 \begin{center}
  \begin{tabular}{|c|c|c|c|}
    \hline
    W boson   & Z boson   &quarks, leptons    & Higgs boson   \\
    \hline
    $\frac{g}{2}v$   &$\frac{\sqrt{g^2+g'^2}}{2} v$  & $y_f \frac{v}{\sqrt{2}}$   & $\sqrt{2\lambda}v$   \\
    \hline
  \end{tabular}
 \end{center}
\end{table}

\section{Naturalness and Physics beyond the Standard Model}
Although the Higgs potential in Eq.\ref{eq1} is very simple 
and sufficient to describe a realistic model of mass generation, 
we think that this is not the final form of the theory but rather an effective 
description of a more fundamental theory.  
It is therefore important to know what is limitation of this description
of the Higgs sector.    

In renormalizable quantum field theories, the form of Lagrangian
is specified by requirement for renormalizability. In the case of the
Higgs potential, quadratic and quartic terms are only renormalizable
interactions. We can then consider two kinds of corrections to the potential. 
One is a calculable higher order correction within the Standard Model.
For instance the correction from the top Yukawa coupling constant can be 
evaluated up to a desired accuracy applying renormalization procedure of
field theory. Another type of corrections comes from outside of 
the present model, presumably from physics at some high energy scale. 
We cannot really compute these corrections 
until we know the more fundamental theory. In this sense, the present 
theory is considered as an effective theory below some cutoff 
energy scale $\Lambda$. 

Although the effective theory cannot include all physical effects, it is
still useful because unknown correction is expected to be suppressed 
by $(E/\Lambda)^2$ where $E$ is a typical energy scale under consideration. 
Therefore, as long as the cutoff scale is somewhat larger than $E$,
the theory can make fairly accurate predictions.  
For example, the correction is $0(10^{-4})$ when the cutoff 
scale is around 10 TeV for physical processes in the 100 GeV range. 
If the theory is valid up to the Planck scale $(\sim 10^{19}$ GeV) where
the gravity interaction becomes as strong as the other gauge interactions,
the correction becomes extremely small. In this way,
an effective theory is useful description as long as we restrict ourself
to the energy regime below the cutoff scale.

Once we take a point of view that the Higgs sector of the Standard Model is
an effective description of a more complete theory below $\Lambda$, 
naturalness with regard to parameter fine-tuning 
becomes a serious problem. In particular, the quadratic divergence of the 
Higgs mass radiative correction is problematic, and this has been one 
of main motivations to introduce various models beyond the Standard Model.

In the Higgs potential in Eq.\ref{eq1} the only mass parameter is $\mu^2$. 
At the tree level, this parameter is related to the vacuum expectation value 
$v$ by $\mu^2=\lambda v^2$ where $v$ is known to be about 246 GeV.
($v=(\sqrt{2} G_F)^{-\frac{1}{2}}$, 
where $G_F$ is the Fermi constant representing
the coupling constant of the weak interaction.)
If we include the radiative correction, $\mu^2$ becomes a sum of two 
contributions $\mu_0^2 + \delta \mu^2$ where $\mu_0^2$
is a bare mass term and $\delta \mu^2$ is the radiative correction. In the 
Standard Model, the top quark and gauge boson loop corrections are 
important and $\delta \mu^2$ from these sources are represented 
by a sum of terms  of a form $C_i \frac{g_i^2}{(4\pi)^2}\Lambda^2$
where $g_i$ is the top Yukawa coupling constant, or $U(1)$ or $SU(2)$ gauge
coupling constant and $C_i$ are $O(1)$ coefficients. Since the radiative
correction depends on the cutoff scale quadratically, the fine-tuning 
between the bare mass term and the radiative correction is necessary 
if the cutoff scale is much larger than 1 TeV. Roughly speaking, the 
fine-tuning at 1\% level is necessary for $\Lambda=10$ TeV. If the cutoff
scale is close to the Planck scale, the degree of the fine-tuning is enormous:
A tuning of one out of $10^{32}$ is required. This is the naturalness problem
of the Standard Model, and sometime also called the hierarchy problem.
This problem suggests that the description of the Higgs sector by the 
simple potential in Eq. \ref{eq1} is not very satisfactory, and probably 
will be replaced by a more fundamental form at a higher energy scale. 

Since the problem arises from the quadratic divergence in the renormalization
of the Higgs mass terms, proposed solutions involve cancellation 
mechanism of such divergence. Supersymmetry\cite{Wess:1974tw}
is a unique symmetry that 
guarantees complete cancellation of the quadratic divergence in 
scalar field mass terms.
This is a new symmetry between bosons and fermions and the cancellation occurs
between loop diagrams of bosons and fermions.  Particle
physics models based on supersymmetry such as supersymmetric grand unified 
theory (SUSY GUT) have been proposed and studied 
since early 1980's as a possible way out of the hierarchy 
problem \cite{Sakai:1981gr}. 
In a realistic model of a supersymmetric extension of the Standard Model,
we need to introduce new particles connected by supersymmetry 
to ordinary quarks, leptons, gauge bosons, and Higgs particles.   
In 1990's, precision studies on the Z boson were preformed at LEP 
and SLC experiments, and it was pointed out that three precisely measured 
coupling constants are consistent with the prediction of SUSY GUT,
although the gauge coupling unification fails badly without
supersymmetric partner 
particles \cite{Langacker:1991an}. 
Supersymmetry is also an essential 
ingredient of the superstring theory, a potential unified theory including
gravity and gauge interactions. In this way, the supersymmetric model has
become a promising candidate beyond the Standard Model. If supersymmetry
realizes at or just above the TeV scale, it can provide
a consistent and unified picture of the particle physics
from the weak scale to the Planck scale.

An opposite idea for solution of the naturalness problem is 
considering that the cutoff scale is close to the electroweak scale. 
In particular, the Higgs field is considered to be a composite state of more 
fundamental objects at a relatively low energy scale.  
The simplest form of this model is called the technicolor model 
\cite{Susskind:1978ms} 
proposed in late 1970's, in which the cutoff scale is about 1 TeV. 
The technicolor model is however strongly constrained from precision
tests of electroweak theory later at the LEP and SLC experiments 
\cite{Peskin:1990zt}, 
but there have been continuous attempts to construct a phenomenologically 
viable model of a composite Higgs field. Little Higgs 
models\cite{Arkani-Hamed:2002qx}
are a recent proposal on this line, where the physical Higgs boson is 
dynamically formed by a new strong interaction around 10 TeV. 
An interesting feature of this model is that the quadratic divergence 
of the Higgs boson mass term is canceled by loop corrections due to 
new gauge bosons and a heavy partner of the top quark at one loop level.
In this way the hierarchy problem between the 
electroweak scale and 10 TeV is nicely solved.       
  
In addition to supersymmetry and little Higgs models, there have been many
proposals for TeV scale physics. Motivations for many of them 
are solving the naturalness problem of the Standard Model or 
explaining the large (apparent) hierarchy between the weak scale and 
the gravity scale. Examples are models with large extra 
dimensions \cite{Arkani-Hamed:1998rs}, 
models with warped 
extra-dimensions \cite{Randall:1999ee},  
the Higgsless model\cite{Csaki:2003dt},
the twin Higgs model\cite{Chacko:2005pe}, 
and the inert Higgs model\cite{Barbieri:2006dq}, 
the split-supersymmetry model\cite{Arkani-Hamed:2004fb}, 
etc. All of these proposals involve some characteristic signals 
around a TeV region. These signals are important to choose a correct model 
at the TeV scale and clarify the mechanism of the electroweak 
symmetry breaking.

\section{Experimental Prospects of Higgs Physics}
Higgs physics is expected to be the center of the particle physics in coming 
years starting from the commissioning of the CERN LHC experiment. 
The first step will be a discovery of a new particle which is a 
candidate of the Higgs boson.
We then study its properties in detail and compare them with the prediction
of the Standard Model Higgs boson. We may be able to confirm that the
discovered particle is the Higgs boson responsible for the mass 
generation for elementary particles. Another possibility 
would be to find some deviation from the Standard Model Higgs boson. 
Deviation could be something like small difference of production cross section
and decay branching ratios from the Standard Model predictions, 
or more drastic new signals such as discovery of
several Higgs states. At the same time, we may also find other new particles 
predicted in extensions of the Standard Model, for example supersymmetric 
particles in the supersymmetric model or the heavy gauge bosons and the 
top partner in the little Higgs model. In order to accomplish these goals
we probably need several steps in collider experiments including LHC
and ILC experiments and possible upgrades for these facilities. 
  
If we restrict ourselves to the Higgs boson in the Standard Model, 
all physical properties are determined by one parameter, the Higgs 
boson mass. Present experimental lower bound for the mass of 
the Standard Model Higgs boson is 114.4 GeV at the 95\% confidence level,
set by the direct Higgs boson search at LEP \cite{Barate:2003sz}. 
It is remarkable that we can also draw an upper bound 
from a global fit of electroweak precision data.
Although a heavy Higgs boson means a large self-coupling $\lambda$, 
we have not seen any evidence of such a large coupling
in physical observables related to $Z$ and $W$ gauge boson processes. 
The upper limit of the Standard Model Higgs boson is 166 GeV 
at the 95\% confidence level \cite{Alcaraz:2006mx}.
This implies that a relatively light Higgs boson is favored. If the Higgs
boson turns out to be heaver than 200 GeV, we would expect some additional 
new particles that have significant couplings to gauge bosons.

\begin{figure}[tb]
\begin{center}
\includegraphics[width=8cm,angle=0]{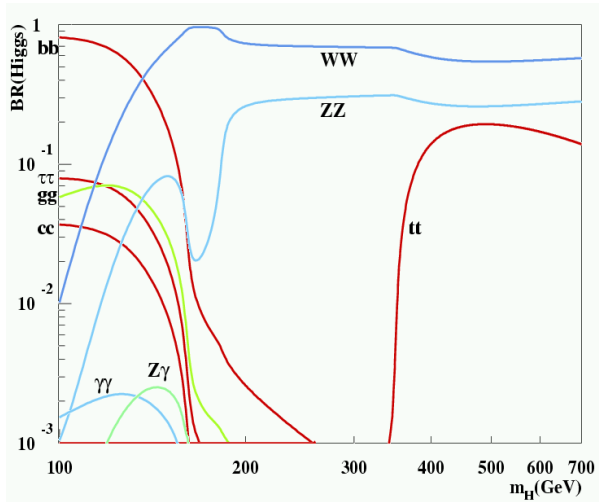}
\end{center}
\caption{Decay branching ratio of the Standard Model Higgs boson as a
function of its mass. $c$, $b$, $t$, $\tau$ represent charm, bottom, top
quarks and tau lepton. $\gamma$, $g$ , $W$, $Z$ are photon, gluon, $W$ 
and $Z$ bosons. }
\label{f1}
\end{figure}

The decay branching ratios of the Higgs boson depends strongly on 
the Higgs boson mass, and therefore the discovery strategy for the 
Higgs boson at LHC differs for light and heavy Higgs bosons. The
branching ratios for the Standard Model Higgs boson is shown in figure 
\ref{f1}. Since the Higgs boson couples more strongly to a heaver particle,
it tends to decay to heaver particles as long as kinematically allowed.
For instance, the Higgs boson mostly decays into two gauge bosons 
if the Higgs boson mass is larger than 200 GeV, whereas the 
bottom and anti-bottom pair is the main decay mode for its mass less than
140 GeV. For this mass range, the Higgs boson search at LHC relies 
on other decay modes such as the loop-induced two photon decay mode,
because two bottom modes are hidden by overwhelming QCD background 
processes. Detail simulation studies on the Higgs discovery at LHC
have been performed, and it is shown that the Higgs boson can be found 
at LHC experiments within a few years for the entire mass region as long as
the production and decay properties are similar to the Standard Model
Higgs boson\cite{atlas:1999, cms:2006}. 
Furthermore, information on the Higgs couplings is obtained 
with a higher luminosity.
Estimated precision for coupling ratios are typically
0(10)\% \cite{Duhrssen:2004cv}.

\begin{figure}[tb]
\begin{center}
\includegraphics{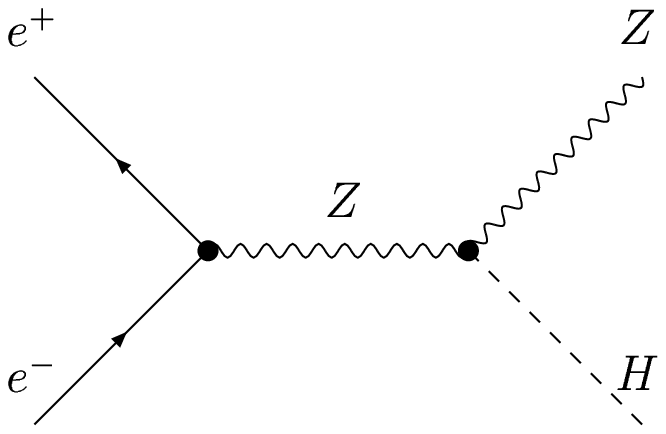}
\hspace{1cm}
\includegraphics{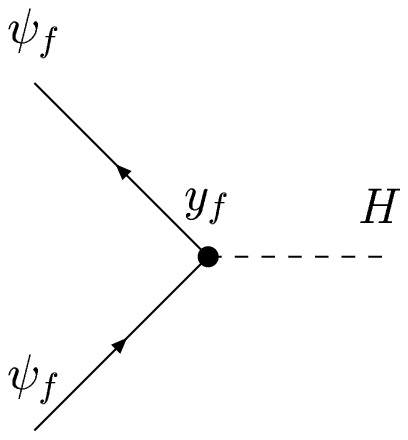}
\end{center}
\caption{Production process of the Higgs boson at ILC (left) and
Higgs boson decay to fermions (right).}
\label{f2}
\end{figure}

\begin{figure}[h] \centering
    \includegraphics[width=8cm,angle=0]{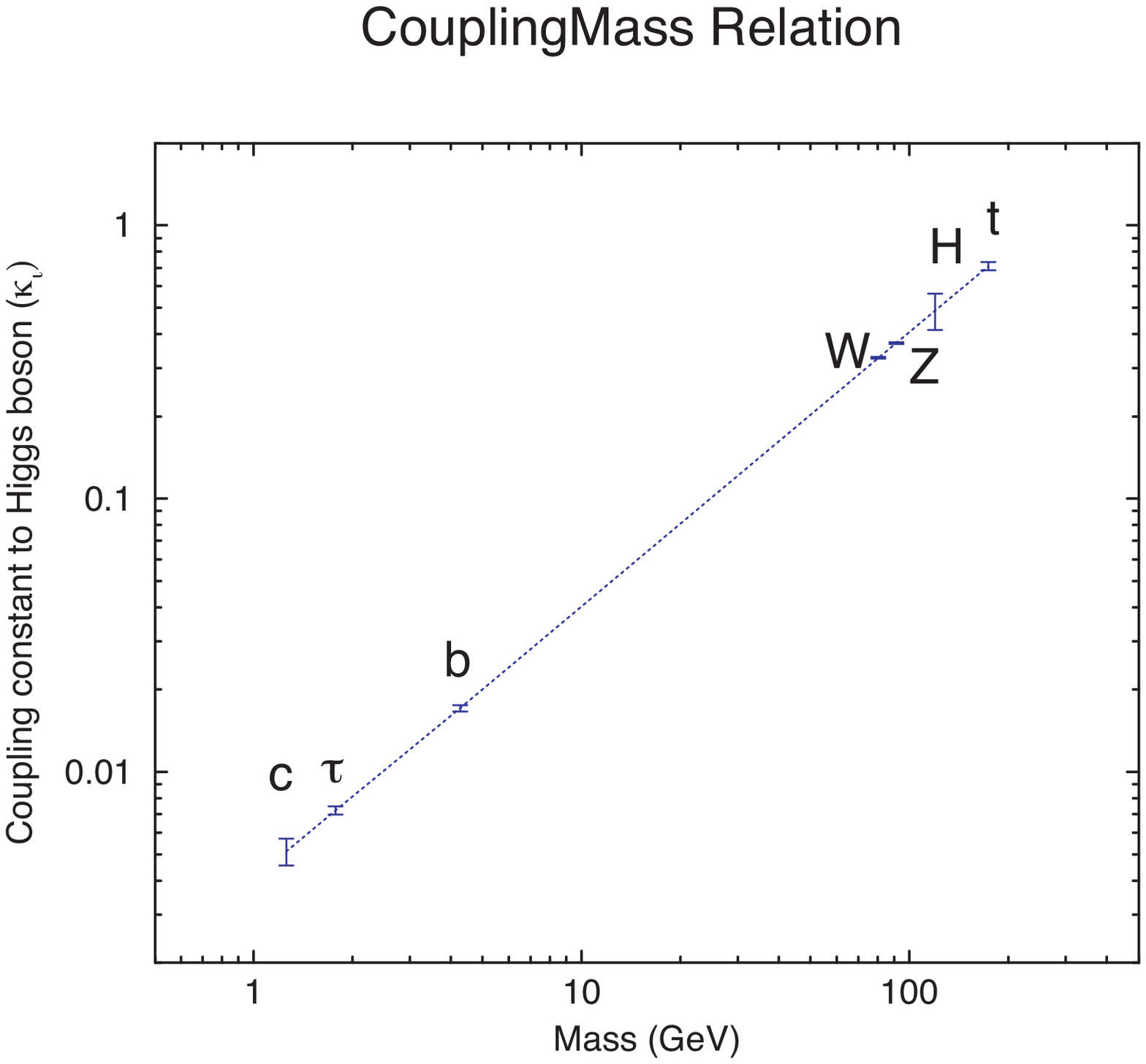}
    \caption{Precision of the coupling-constant determination for
    various particles at ILC with the integrated luminosity of
    500 $fb^{-1}$. The Higgs boson mass is taken to be 120 GeV.
    For charm, tau, bottom, W, and Z coupling measurement, $\sqrt{s}$=300 GeV
    is assumed. $\sqrt{s}=500$ GeV (700 GeV) is taken for the triple
    Higgs $(t\bar{t} H)$ coupling measurement\cite{:2003mg}.}
\label{f3} 
\end{figure}

ILC is a future electron-positron linear collider project 
proposed in the international framework \cite{ILC}. 
One aspect of this facility is a Higgs factory. 
For instance, the number of produced Higgs bosons can 
be $0(10^5)$ in the first stage of experiments with the 
center-of-mass collider energy of 500 GeV. Under clean environment of 
the $e^+e^-$ collider, precise determinations on the mass, quantum numbers, 
and coupling constants of the
Higgs boson are possible. Typical production and decay processes 
are shown figures \ref{f2}.  Precision of the coupling constant
determination reaches a few \% level for Higgs-$WW$, Higgs-$ZZ$,
and Higgs-$b \bar{b}$ couplings for the case of 
a relatively light Higgs boson. We can also measure the Higgs self-coupling 
from the double Higgs boson production process and the top Yukawa coupling
from the Higgs-$t \bar{t}$ production. Figure \ref{f3}
shows precision of the Higgs coupling constant determination for 
various particles at ILC. The proportionality between coupling constants
and particle masses is a characteristic feature of the one Higgs doublet
model where the particle mass formulas involve only one vacuum expectation 
value. An important feature of ILC experiments 
is that absolute values of these coupling constants can be 
determined in a model-independent way. This is crucial in 
establishing the mass generation mechanism for elementary particles.   

The precise determination of the Higgs coupling constants is also useful
to explore physics beyond the Standard Model. In some case, the Higgs boson
coupling is modified from the Standard Model.

\begin{itemize}
\item 
The Higgs sector of supersymmetric models is different from
the Standard Model. In any realistic supersymmetric model, 
the Higgs sector contains at lease two sets of doublet fields. 
In the minimal supersymmetric standard model (MSSM), in particular, 
the Higgs sector is a two Higgs doublet model. Furthermore,
there is a rather strict theoretical upper bound for the
the lightest neutral Higgs boson\cite{Okada:1990vk}, 
which is about 130 GeV.
Since this light boson plays a role of the usual Higgs 
particle, this particle may be the only Higgs particle discovered
at LHC. In such case, the branching ratio 
measurement for the lightest neutral Higgs boson is useful 
to obtain information on
the masses of heavy Higgs bosons 
\cite{Abe:2001gc, Aguilar-Saavedra:2001rg, Weiglein:2004hn}. 
In particular, the tau and 
bottom coupling constants show sizable enhancement if the heavy
Higgs boson exists below 600 GeV. The ratio like 
$B(H \to WW)/B(H \to \tau \tau)$ is useful to determine 
the heavy Higgs mass scale indirectly.   
\item In models with extra dimensions, there appears a scalar field called
Radion, corresponding to the size of the extra space dimension. 
Since Radion is a neutral scalar field, it can mix with the
Higgs field. It is pointed out that Radion-Higgs mixing in the 
warped extra dimension model could reduce the magnitude of 
Yukawa coupling constants and $WWH$ and $ZZH$
constant in a universal way \cite{Hewett:2002nk}. 
In order to observe such effects,
absolute coupling measurements at ILC are necessary.    
\item The two-gluon width of the Higgs boson is generated by loop
diagrams, so that it can be a probe to virtual effects of 
new particles. The same is true for the two-photon width, 
the measurement of which is improved at the photon-photon collider 
option of ILC \cite{Badelek:2001xb}. 
There are many new physics models where such 
loop effects are sizable.  
\item Explaining the baryon number of the Universe is one of most 
outstanding questions for particle physics in connection 
with cosmology \cite{Riotto:1999yt}. 
One possibility is the electroweak baryogenesis scenario, in which the 
baryon number was generated at the electroweak phase transition. 
For a successful electroweak baryogenesis, the Higgs sector has to 
be extended from that of the minimal Standard Model to realize a strong 
first-order phase transition. The change of the Higgs potential
can lead to observable effects in the triple Higgs coupling 
measurement \cite{Grojean:2004xa,Kanemura:2004ch}.   
\end{itemize}

As we can see above examples, observations of new physics effects 
require precise determination of coupling constants. This will be an
important goal of the future ILC experiment.

\section{Conclusions}
The Higgs sector is an unknown part of the particle physics model. 
Although a simple potential is assumed in the Standard Model, 
this description is supposed to be valid below some cutoff scale, 
beyond which the theory of the electroweak symmetry breaking 
takes in a more fundamental form. If the cutoff scale is as low as 1 TeV, some direct signals on new physics is likely to appear at LHC. 
If the cutoff scale is much larger, the fine-tuning of the Higgs boson mass 
term becomes a serious problem. Proposed solutions to this problem such 
as supersymmetry or little Higgs models also predict new physics signals 
at the TeV scale. These signals are targets of future collider experiments
starting from LHC.      

Experimental prospects for the Higgs physics are quite bright. The Higgs 
particle can be found and studied at LHC.  At the proposed ILC, precise 
information on coupling constants between the Higgs boson and other particles 
will be obtained. These measurements are an essential step to establish 
the mass generation mechanism. 
At the same time, the precision measurement may reveal evidence of
new force and/or new symmetry because these new physics is most probably
related to the physics of electroweak symmetry breaking, i.e. the 
Higgs sector. In this way, the Higgs particle 
will play a special role in determining the future direction of the 
particle physics.

\section*{Acknowledgment}
This work was supported in part by the Grant-in-Aid for Science
Research, Ministry of Education, Culture, Sports, Science and 
Technology, Nos. 16081211 and 17540286.


\begin{thebibliography}{99} %% The number "99" means that this list has more than nine items.
\bibitem{Nambu:1961tp}
  Y.~Nambu and G.~Jona-Lasinio,
  %``Dynamical model of elementary particles based on an analogy with
  %superconductivity. I,''
  Phys.\ Rev.\  {\bf 122} (1961) 345.
\bibitem{Goldstone:1961eq}
  J.~Goldstone,
  %``Field Theories With Superconductor Solutions,''
  Nuovo Cim.\  {\bf 19} (1961) 154.
\bibitem{Higgs:1964ia}
  P.~W.~Higgs,
  %``Broken symmetries, massless particles and gauge fields,''
  Phys.\ Lett.\  {\bf 12} (1964) 132;
%``BROKEN SYMMETRIES AND THE MASSES OF GAUGE BOSONS,''
  Phys.\ Rev.\ Lett.\  {\bf 13} (1964) 508;
%``Spontaneous Symmetry Breakdown Without Massless Bosons,''
  Phys.\ Rev.\  {\bf 145} (1966) 1156.  
\bibitem{Weinberg:1967tq}
  S.~Weinberg,
  %``A Model Of Leptons,''
  Phys.\ Rev.\ Lett.\  {\bf 19} (1967) 1264.
\bibitem{Lee:1956qn}
  T.~D.~Lee and C.~N.~Yang,
  %``Question Of Parity Conservation In Weak Interactions,''
  Phys.\ Rev.\  {\bf 104} (1956) 254.
\bibitem{Wess:1974tw}
  J.~Wess and B.~Zumino,
  %``Supergauge Transformations in Four-Dimensions,''
  Nucl.\ Phys.\  B {\bf 70} (1974) 39.
\bibitem{Sakai:1981gr}
  N.~Sakai,
  %``Naturalness In Supersymmetric Guts,''
  Z.\ Phys.\  C {\bf 11} (1981) 153;
  S.~Dimopoulos and H.~Georgi,
  %``Softly Broken Supersymmetry And SU(5),''
  Nucl.\ Phys.\  B {\bf 193} (1981) 150.
\bibitem{Langacker:1991an}
  P.~Langacker and M.~x.~Luo,
  %``Implications of precision electroweak experiments for M(t), rho(0),
  %sin**2-Theta(W) and grand unification,''
  Phys.\ Rev.\  D {\bf 44} (1991) 817;
 U.~Amaldi, W.~de Boer and H.~Furstenau,
  %``Comparison of grand unified theories with electroweak and strong coupling
  %constants measured at LEP,''
  Phys.\ Lett.\  B {\bf 260} (1991) 447.
\bibitem{Susskind:1978ms}
  L.~Susskind,
  %``Dynamics Of Spontaneous Symmetry Breaking In The Weinberg-Salam Theory,''
  Phys.\ Rev.\  D {\bf 20} (1979) 2619.
\bibitem{Peskin:1990zt}
  M.~E.~Peskin and T.~Takeuchi,
  %``A New constraint on a strongly interacting Higgs sector,''
  Phys.\ Rev.\ Lett.\  {\bf 65} (1990) 964;
  M.~Golden and L.~Randall,
  %``RADIATIVE CORRECTIONS TO ELECTROWEAK PARAMETERS IN TECHNICOLOR THEORIES,''
  Nucl.\ Phys.\  B {\bf 361} (1991) 3;
 B.~Holdom and J.~Terning,
  %``Large corrections to electroweak parameters in technicolor theories,''
  Phys.\ Lett.\  B {\bf 247} (1990) 88.
\bibitem{Arkani-Hamed:2002qx}
  N.~Arkani-Hamed, A.~G.~Cohen, E.~Katz, A.~E.~Nelson, T.~Gregoire and J.~G.~Wacker,
  %``The minimal moose for a little Higgs,''
  JHEP {\bf 0208} (2002) 021;.
 N.~Arkani-Hamed, A.~G.~Cohen, E.~Katz and A.~E.~Nelson,
  %``The littlest Higgs,''
  JHEP {\bf 0207} (2002) 034.
\bibitem{Arkani-Hamed:1998rs}
  N.~Arkani-Hamed, S.~Dimopoulos and G.~R.~Dvali,
  %``The hierarchy problem and new dimensions at a millimeter,''
  Phys.\ Lett.\  B {\bf 429} (1998) 263;
 I.~Antoniadis, N.~Arkani-Hamed, S.~Dimopoulos and G.~R.~Dvali,
  %``New dimensions at a millimeter to a Fermi and superstrings at a TeV,''
  Phys.\ Lett.\  B {\bf 436} (1998) 257;
  N.~Arkani-Hamed, S.~Dimopoulos and G.~R.~Dvali,
  %``Phenomenology, astrophysics and cosmology of theories with  sub-millimeter
  %dimensions and TeV scale quantum gravity,''
  Phys.\ Rev.\  D {\bf 59} (1999) 086004.
\bibitem{Randall:1999ee}
  L.~Randall and R.~Sundrum,
  %``A large mass hierarchy from a small extra dimension,''
  Phys.\ Rev.\ Lett.\  {\bf 83} (1999) 3370.
\bibitem{Csaki:2003dt}
  C.~Csaki, C.~Grojean, H.~Murayama, L.~Pilo and J.~Terning,
  %``Gauge theories on an interval: Unitarity without a Higgs,''
  Phys.\ Rev.\  D {\bf 69} (2004) 055006;
  C.~Csaki, C.~Grojean, L.~Pilo and J.~Terning,
  %``Towards a realistic model of Higgsless electroweak symmetry breaking,''
  Phys.\ Rev.\ Lett.\  {\bf 92} (2004) 101802.
\bibitem{Chacko:2005pe}
  Z.~Chacko, H.~S.~Goh and R.~Harnik,
  %``The twin Higgs: Natural electroweak breaking from mirror symmetry,''
  Phys.\ Rev.\ Lett.\  {\bf 96} (2006) 231802. 
\bibitem{Barbieri:2006dq}
  R.~Barbieri, L.~J.~Hall and V.~S.~Rychkov,
  %``Improved naturalness with a heavy Higgs: An alternative road to LHC
  %physics,''
  Phys.\ Rev.\  D {\bf 74} (2006) 015007.
\bibitem{Arkani-Hamed:2004fb}
  N.~Arkani-Hamed and S.~Dimopoulos,
  %``Supersymmetric unification without low energy supersymmetry and  signatures
  %for fine-tuning at the LHC,''
  JHEP {\bf 0506} (2005) 073.
\bibitem{Barate:2003sz}
  R.~Barate {\it et al.}  [LEP Working Group for Higgs boson searches],
  %``Search for the standard model Higgs boson at LEP,''
  Phys.\ Lett.\  B {\bf 565} (2003) 61.  
\bibitem{Alcaraz:2006mx}
  J.~Alcaraz {\it et al.}  The LEP Collaborations and the LEP electroweak
  working group,
  %``A combination of preliminary electroweak measurements and constraints on
  %the standard model,''
  arXiv:hep-ex/0612034.
\bibitem{atlas:1999}
 ATLAS Collaboration, ATLAS Physics Technical Report,
 CERN-LHCC-99-14 and CERN-LHCC-99-15.
\bibitem{cms:2006}
 CMS Collaboration, CMS Physics TDR,
 CERN/LHCC/2006-021. 
\bibitem{Duhrssen:2004cv}
  M.~Duhrssen, S.~Heinemeyer, H.~Logan, D.~Rainwater, G.~Weiglein and D.~Zeppenfeld,
  %``Extracting Higgs boson couplings from LHC data,''
  Phys.\ Rev.\  D {\bf 70} (2004) 113009.
\bibitem{ILC}
International Linear Collider home page,
http://www.linearcollider.org/
\bibitem{:2003mg}
  ``GLC project: Linear collider for TeV physics,''
  KEK-REPORT-2003-7.
\bibitem{Okada:1990vk}
  Y.~Okada, M.~Yamaguchi and T.~Yanagida,
  %``Upper bound of the lightest Higgs boson mass in the minimal supersymmetric
  %standard model,''
  Prog.\ Theor.\ Phys.\  {\bf 85} (1991) 1;
 J.~R.~Ellis, G.~Ridolfi and F.~Zwirner,
  %``Radiative corrections to the masses of supersymmetric Higgs bosons,''
  Phys.\ Lett.\  B {\bf 257} (1991) 83;
  H.~E.~Haber and R.~Hempfling,
  %``Can the mass of the lightest Higgs boson of the minimal supersymmetric
  %model be larger than m(Z)?,''
  Phys.\ Rev.\ Lett.\  {\bf 66} (1991) 1815.
\bibitem{Abe:2001gc}
  K.~Abe {\it et al.}  [ACFA Linear Collider Working Group],
  ``Particle physics experiments at JLC,''
  arXiv:hep-ph/0109166.
\bibitem{Aguilar-Saavedra:2001rg}
  J.~A.~Aguilar-Saavedra {\it et al.}  [ECFA/DESY LC Physics Working Group],
  `TESLA Technical Design Report Part III: Physics at an e+e- Linear
  Collider,''
  arXiv:hep-ph/0106315.
\bibitem{Weiglein:2004hn}
  G.~Weiglein {\it et al.}  [LHC/LC Study Group],
  ``Physics interplay of the LHC and the ILC,''
  Phys.\ Rept.\  {\bf 426} (2006) 47
  [arXiv:hep-ph/0410364].
\bibitem{Hewett:2002nk}
  J.~L.~Hewett and T.~G.~Rizzo,
  %``Shifts in the properties of the Higgs boson from radion mixing,''
  JHEP {\bf 0308} (2003) 028;
  D.~Dominici, B.~Grzadkowski, J.~F.~Gunion and M.~Toharia,
  %``The scalar sector of the Randall-Sundrum model,''
  Nucl.\ Phys.\  B {\bf 671} (2003) 243.
\bibitem{Badelek:2001xb}
  B.~Badelek {\it et al.}  [ECFA/DESY Photon Collider Working Group],
  %``TESLA Technical Design Report, Part VI, Chapter 1: Photon collider  at
  %TESLA,''
  Int.\ J.\ Mod.\ Phys.\  A {\bf 19} (2004) 5097.
\bibitem{Riotto:1999yt}
  A.~Riotto and M.~Trodden,
  %``Recent progress in baryogenesis,''
  Ann.\ Rev.\ Nucl.\ Part.\ Sci.\  {\bf 49} (1999) 35;
  M.~Dine and A.~Kusenko,
  %``The origin of the matter-antimatter asymmetry,''
  Rev.\ Mod.\ Phys.\  {\bf 76} (2004) 1;
   W.~Buchmuller, R.~D.~Peccei and T.~Yanagida,
  %``Leptogenesis as the origin of matter,''
  Ann.\ Rev.\ Nucl.\ Part.\ Sci.\  {\bf 55} (2005) 311.
\bibitem{Grojean:2004xa}
  C.~Grojean, G.~Servant and J.~D.~Wells,
  %``First-order electroweak phase transition in the standard model with a  low
  %cutoff,''
  Phys.\ Rev.\  D {\bf 71} (2005) 036001.
\bibitem{Kanemura:2004ch}
  S.~Kanemura, Y.~Okada and E.~Senaha,
  %``Electroweak baryogenesis and quantum corrections to the triple Higgs  boson
  %coupling,''
  Phys.\ Lett.\  B {\bf 606} (2005) 361.
\end{thebibliography}
\end{document}